\documentclass{iopjournal}

\usepackage{amsmath,amssymb}
\usepackage[numbers,sort&compress]{natbib}
\usepackage{booktabs}
\usepackage{graphicx}
\usepackage[table]{xcolor}
\usepackage{orcidlink}
\let\orcid\orcidlink

\renewcommand{\articletype}[1]{{\vspace*{-8mm}\noindent \Large \sf Measurement Science and Technology}
\par\vspace{4mm}\noindent\normalsize #1\par\vspace{4mm}}

\newcommand{\SII}{S_{21}}
\newcommand{\dSII}{\Delta S_{21}}
\newcommand{\Np}{\ensuremath{\,\mathrm{Np}}}
\newcommand{\dB}{\ensuremath{\,\mathrm{dB}}}
\newcommand{\GHz}{\ensuremath{\,\mathrm{GHz}}}

\newcommand{\Rs}{R_\mathrm{s}}
\newcommand{\Zo}{Z_0}

\newcommand{\eeff}{\varepsilon_\mathrm{eff}}
\newcommand{\Nsq}{N_{\square}}
\newcommand{\nptunerurl}{https://github.com/scottdietrich/NpTuner}

\begin{document}

\articletype{Paper}

\title{Sensitivity-optimal coplanar waveguide design for broadband magnetic resonance spectroscopy: a Beer--Lambert framework}

\author{Scott Dietrich$^{1,*}$\orcid{0000-0001-6405-353X}, and Artur Solodovnyk$^1$\orcid{0000-0002-4115-5020}}

\affil{$^1$ Department of Physics, Villanova University, Villanova, PA 19085, USA}
\affil{$^*$ Author to whom any correspondence should be addressed.}
\email{scott.dietrich@villanova.edu}


\begin{abstract}
Coplanar waveguide (CPW) transmission spectroscopy is used to probe spin dynamics, ferromagnetic resonance, and complex conductivity across a wide range of materials, yet no systematic framework connects waveguide geometry to measurement sensitivity when sample volume and concentration are fixed. We show that the shared geometric scaling of sample coupling and conductor loss maps CPW design onto the Beer--Lambert optimization problem of optical spectrophotometry, reducing it to a universal one-parameter problem whose solution depends only on sample geometry and the dominant noise source---not on sample properties, operating frequency, or system losses. The framework predicts a near-universal $1\,\mathrm{Np}$ optimum across the full range of sample thicknesses and design geometries. Benchmarking against seven published broadband FMR instruments reveals two fabrication-delimited classes: PCB-milled designs are bounded by a ceiling imposed by their minimum slot width, while photolithographic designs approach the additive-noise optimum. For large-area PCB samples a meander geometry offers a direct path to near-optimal sensitivity without interferometric compensation; for sub-millimeter samples, a single lithographic straight pass suffices.
\end{abstract}

\section{Introduction}
\label{sec:intro}

Microwave-frequency spectroscopy has become a routine tool for probing spin dynamics, ferromagnetic resonance, and complex conductivity across a wide range of materials. Broadband techniques operating in the 0.1--20~GHz range use a coplanar waveguide (CPW) as both the microwave delivery structure and the transduction element: sample-induced changes in the transmitted signal $\SII(f)$ encode the magnetic or electric response of the material under study. A vector network analyzer (VNA) driving a CPW delivers continuous frequency coverage with high dynamic range, making it well suited for mapping dispersion relations, identifying phase transitions, and characterizing anisotropy: a single magnetic-field sweep returns the complete $\SII(f, B_0)$ map across the operating frequency range~\cite{Bilzer2007} -- a capability that cavity-based electron paramagnetic resonance (EPR) and pulsed time-domain methods cannot match without either sacrificing bandwidth or adding substantial instrumentation complexity \cite{Poole1996,Schweiger2001}.

Classical CPW design for 50~$\Omega$ impedance matching on printed-circuit-board (PCB) substrates is a well-documented standard practice \cite{Simons2001,Pozar2012}. Standard guidelines specify the ratio of conductor width $W$ to slot width $S$ for a target characteristic impedance, and tabulate dielectric and conductor losses for common substrate materials. These guidelines assume, implicitly or explicitly, that the sample either fills the waveguide environment or extends uniformly over the full CPW surface. Neither assumption addresses the fundamental sensitivity question: given a fixed sample volume and concentration that cannot be increased, what CPW geometry maximises the signal-to-noise ratio? This question has a general answer rooted in transmission spectrophotometry. In optical Beer--Lambert spectroscopy, the signal-to-noise ratio of an absorbance measurement is maximized at a specific path-length-dependent attenuation that depends on the dominant noise source \cite{Rothman1975,Lothian1949}. An identical trade-off governs CPW transmission: signal coupling grows with conductor length and narrowness, but so does the background attenuation that the signal must overcome. The framework applies at any sample scale and becomes the dominant design constraint whenever sample volume or concentration is fixed. What the optical framework does not provide is the translation from an optimal attenuation target to a specific waveguide geometry. This translation requires filling-factor, conductor-loss, and meander analyses specific to the CPW geometry.

The same Beer--Lambert structure applies equally to magnetic-field-coupled measurements (EPR, FMR, spin-wave spectroscopy) and electric-field-coupled measurements
(complex conductivity spectroscopy~\cite{Engel1993,Dietrich2025});
the present treatment focuses on the magnetic case though it can easily
be extended to the latter. The sensitivity of CPW-based magnetic resonance was first analyzed quantitatively by Narkowicz \emph{et al.}~\cite{Narkowicz2005,Narkowicz2008}, who introduced the filling factor concept for small samples on CPW structures. Clauss \textit{et al.}~\cite{Clauss2013}, Ghirri \textit{et al.}~\cite{Ghirri2016}, and Boudalis \textit{et al.}~\cite{Boudalis2021} developed broadband FMR and EPR protocols using CPW excitation, establishing the technique across a range of materials. The foundational conductor loss models of Pucel \emph{et al.}~\cite{Pucel1968} and the thin-film extensions of Heinrich~\cite{Heinrich1993} provide the analytical basis for loss prediction in planar waveguides. What is missing is a unified treatment of CPW geometry optimisation for fixed-volume samples that incorporates filling factor, coverage geometry, thin-film metallization, and substrate selection into a single framework.

This paper develops such a framework. Section~\ref{sec:methods} sets out the geometric and loss scaling laws (filling factor, conductor loss, dielectric loss) that reduce the optimization to a single design variable $W$ at fixed $Z_0$. Section~\ref{sec:results} substitutes these into the SNR figure of merit and derives the optimal operating point and its design implications. Section~\ref{sec:discussion} extends the analysis to the noise-mixing parameter $\varphi$, thin-film metallization, and benchmarking against published broadband FMR instruments. An interactive design tool implementing the full analytical model is described in section~\ref{sec:tool}.

\section{Methods}
\label{sec:methods}

\subsection{Measurement geometry and the Beer--Lambert analogy}
\label{sec:geometry}

\begin{figure}[t]
\centering
\includegraphics[width=\linewidth]{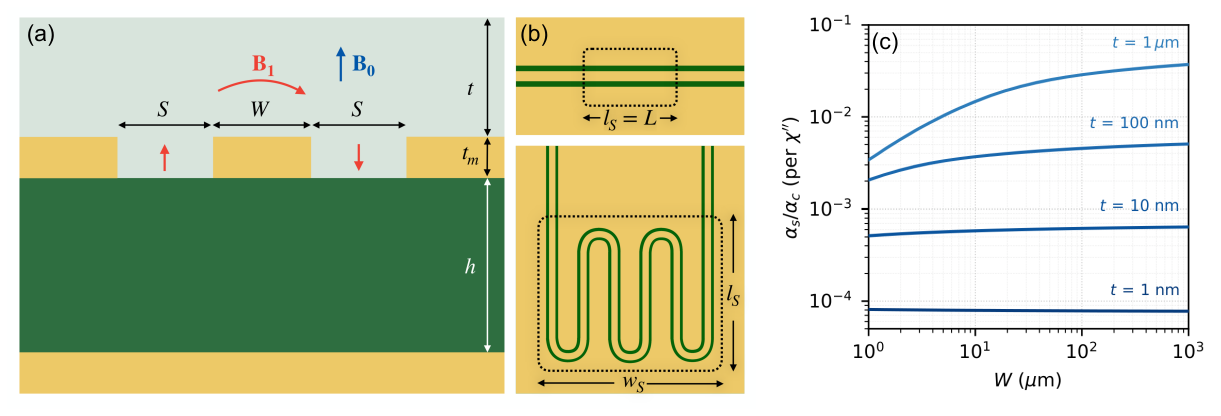}
\caption{CPW geometry and sample coverage. (a)~Cross-section: center conductor (width $W$), slots (width $S$), ground planes, dielectric substrate ($\varepsilon_r$, $h$), and sample layer (thickness $t$). The quasi-static RF magnetic field $\mathbf{B}_1$ is concentrated above the slots and center conductor. (b)~Plan-view schematics of the two design geometries discussed in section~\ref{sec:results}: a single straight pass with interaction length $l_s = L$ (top), and a meander routing multiple parallel passes through a fixed sample area $w_s \times l_s$ with total path length $L$ (bottom). (c)~Coupling-to-loss ratio $\alpha_\mathrm{s}/\alpha_\mathrm{c}$ (per unit $\chi''$) versus conductor width $W$ for four sample thicknesses spanning three decades, computed from the full conformal-mapping integral at fixed $Z_0 = 50\,\Omega$. In the thin-sample regime ($t/W \ll 1$) the ratio is approximately linear in $t$ and nearly independent of $W$, reflecting the shared $1/W$ scaling of $\alpha_\mathrm{s}$ and $\alpha_\mathrm{c}$. Departures appear for thicker samples at small $W$, where $t/W \to 1$, the filling factor $\eta$ saturates, and the thin-sample approximation breaks down.}
\label{fig:geometry}
\end{figure}

The measurement geometry is shown schematically in figure~\ref{fig:geometry}. A coplanar waveguide consists of a center conductor of width $W$ separated from two semi-infinite ground planes by slots of width $S$, all patterned on a dielectric substrate of relative permittivity $\varepsilon_r$ and thickness $h$. The ground-to-ground distance is $G = W + 2S$. A sample of thickness $t$ is placed on the CPW surface in direct contact with the conductors or separated by a thin spacer layer.

A microwave source injects power at one end of the CPW and a detector measures the complex transmission coefficient $\SII(f)$ at the other end as a function of frequency $f$; a VNA serves both roles in the common implementation. When the sample responds to the applied field or frequency, it absorbs microwave energy and produces a change in the transmitted signal whose frequency dependence encodes the physical parameters of interest. The RF magnetic field amplitude above the center conductor scales as $B_1 \propto 1/W$ at fixed input power~\cite{Hoult1976,Narkowicz2005}, so wider conductors produce weaker excitation fields.  The optimization framework developed here assumes operation in the linear regime, where the sample response is proportional to $B_1$.  This requires $B_1$ to remain below the saturation threshold (power broadening in EPR, spin-wave instabilities in FMR); this upper bound sets a practical lower bound on conductor width that is separate from the conductor-loss optimization.

For a given characteristic impedance $\Zo$, the ratio $S/W$ is fixed by the substrate permittivity through the conformal-mapping relation \cite{Simons2001,Wen1969}:
\begin{equation}
  \Zo = \frac{30\pi}{\sqrt{\eeff}} \frac{K(k')}{K(k)},
  \label{eq:Z0}
\end{equation}
where $K$ is the complete elliptic integral of the first kind, $k = W/(W+2S)$, $k' = \sqrt{1-k^2}$, and $\eeff$ is the effective dielectric constant accounting for the mixed air--substrate field distribution. Throughout this paper we parameterize designs by $W$ and compute $S$ from equation~\ref{eq:Z0}, so that $W$ is the single geometric design variable.

Two attenuation mechanisms determine the transmitted signal: absorption by the sample (attenuation coefficient $\alpha_\mathrm{s}$) and background loss $\alpha$ from ohmic dissipation and dielectric loss in the substrate. The change in transmission due to magnetic absorption is
\begin{equation*}
  \dSII = e^{-\alpha L}\!\left[1 - e^{-\alpha_\mathrm{s} L}\right],
\end{equation*}
where $L$ is the CPW length. For the weakly coupled systems that motivate this work, $\alpha_\mathrm{s} L \ll 1$ the equation linearizes to
\begin{equation*}
  |\dSII| \approx \alpha_\mathrm{s} L \cdot e^{-\alpha L}.
\end{equation*}
This equation has the structure of the Beer--Lambert law: a signal term linear in path length ($\alpha_\mathrm{s} L$) multiplied by an exponential background loss ($e^{-\alpha L}$). Under additive noise (constant noise independent of received power), the signal-to-noise ratio for detecting $\dSII$ is
\begin{equation}
  \mathrm{SNR} \propto \alpha_\mathrm{s} L \cdot e^{-\alpha L}.
  \label{eq:SNR}
\end{equation}
This expression has a maximum at a finite value of $\alpha L$: increasing the line length improves the signal linearly, but the exponential noise penalty eventually dominates. The actual optimum depends on the noise regime of the measurement --- whether additive receiver noise or multiplicative baseline ripple dominates --- and is discussed in section~\ref{sec:phi}. Identifying the optimal $\alpha L$ and translating it into a CPW geometry requires understanding how $\alpha_\mathrm{s}$ and $\alpha$ depend on conductor width, which is the task of sections~\ref{sec:filling} and \ref{sec:loss}.

\subsection{Filling factor and coupling strength}
\label{sec:filling}

When microwave current $I$ flows through the center conductor, it generates an RF magnetic field $\mathbf{B}_1$ strongest above the conductor and slots, decaying with height. For frequencies well below the first transverse resonance, the field distribution is quasi-static and can be obtained from conformal mapping \cite{Simons2001}. Above the center conductor, the field is predominantly in-plane ($B_{1x}$), which drives magnetic resonance transitions when $B_0$ is applied perpendicular to the substrate.

The sample-induced attenuation coefficient is obtained by comparing the power absorbed per unit length to the power flow $P_0 = I^2\Zo/2$ in the line \cite{Narkowicz2005}.  The factor $\sqrt{\eeff}/c$ arises from the phase velocity $v_p = c/\sqrt{\eeff}$ of the quasi-TEM mode: for a TEM line the total stored magnetic energy per unit length is proportional to $P_0/v_p$, converting the field-energy denominator of $\eta$ into a quantity proportional to $I^2$:
\begin{equation}
  \alpha_\mathrm{s}
    = \frac{\pi f \chi'' \sqrt{\eeff}}{c} \, \eta,
  \label{eq:alpha_s}
\end{equation}
where $\chi''$ is the imaginary part of the sample's magnetic susceptibility, and the dimensionless transverse filling factor
\begin{equation}
  \eta = \frac{\displaystyle\int_\mathrm{sample} |B_1|^2 \,\mathrm{d}A}
              {\displaystyle\int_\mathrm{all} |B_1|^2 \,\mathrm{d}A}
  \label{eq:eta_def}
\end{equation}
is the fraction of RF magnetic field energy per unit CPW length residing within the sample cross-section, paralleling the filling factor of cavity EPR \cite{Poole1996} adapted to the open geometry of a transmission line.

The conformal-mapping solution shows that the in-plane RF field decays with a characteristic length set by $G = W + 2S \propto W$ at fixed impedance. For a thin sample ($t \ll W$) deposited on the CPW surface:
\begin{equation}
  \eta \approx \frac{C_\eta \, t}{W},
  \label{eq:eta_scaling}
\end{equation}
where $C_\eta$ is a dimensionless constant of order unity depending on $S/W$ and the sample's lateral extent relative to the CPW features. Narrower conductors confine the RF field closer to the surface, increasing the fraction of field energy interacting with a thin sample. The linear scaling holds for $t/W \lesssim 0.1$; for thicker samples the field varies significantly through the sample depth, $\eta$ saturates, and the $1/W$ scaling of $\alpha_\mathrm{s}$ breaks down. In this regime the full conformal-mapping integral should be used and is available in the design tool (section~\ref{sec:tool}).

Combining equations~\ref{eq:alpha_s} and \ref{eq:eta_scaling}:
\begin{equation}
  \alpha_\mathrm{s} = \frac{\pi f \chi'' \sqrt{\eeff}}{c}\,\eta
         \propto \frac{t}{W}
  \label{eq:alphas_fullcov}
\end{equation}
for a sample covering the full lateral extent of the CPW. At fixed $\Zo$, narrowing $W$ increases $\eta$ but does not change the total stored energy per unit current, so $\alpha_\mathrm{s} \propto t/W$.

\subsection{Loss mechanisms in CPW}
\label{sec:loss}

The background attenuation $\alpha = \alpha_\mathrm{c} + \alpha_\mathrm{d}$ has two physical origins: ohmic (conductor) loss $\alpha_\mathrm{c}$ and dielectric loss $\alpha_\mathrm{d}$ in the substrate.

The conductor loss for a CPW with center conductor width $W$ and slot width $S$ is \cite{Pucel1968,Gupta1996}
\begin{equation}
  \alpha_\mathrm{c} = \frac{\Rs}{4 \Zo [K(k)]^2}
    \left[
      \frac{\pi + \ln(4\pi W/t_\mathrm{m})}{W}
      + \frac{\pi + \ln(4\pi S/t_\mathrm{m})}{S}
    \right],
  \label{eq:alpha_c_full}
\end{equation}
where $\Rs$ is the surface resistance and $t_\mathrm{m}$ is the metal thickness. At fixed impedance (constant $S/W$) the dominant scaling is $\alpha_\mathrm{c} \approx A_\mathrm{c}\Rs/W$, where $A_\mathrm{c}$ varies by less than 20\% over two decades of $W$; the full expression~(\ref{eq:alpha_c_full}) is used in all numerical evaluations. The total conductor loss over length $L$ is
\begin{equation}
  \alpha_\mathrm{c} L = A_\mathrm{c}\,\Rs \times \frac{L}{W},
  \label{eq:Nsq}
\end{equation}
where $L/W$ is the aspect ratio of the conductor strip. The translation of the optimization targets into specific aspect ratios for a given metallization is developed in section~\ref{sec:thinfilm}. For a bulk conductor with conductivity $\sigma$, $\Rs = \sqrt{\pi f \mu_0/\sigma}$ grows as $\sqrt{f}$ with frequency. The behavior of $\Rs$ in the thin-film regime, where metallization thickness is comparable to or less than the skin depth, and its implications for the optimization are discussed in section~\ref{sec:thinfilm}.

Dielectric loss $\alpha_\mathrm{d}$ is independent of $W$ and depends only on substrate material and operating frequency; on fused silica at 10~GHz it contributes less than 2\% of the conductor loss at $u = 1\,\mathrm{Np}$ and is neglected throughout. Substrate permittivity does enter through the impedance condition (equation~\ref{eq:Z0}), which sets the $S/W$ ratio and thereby the minimum achievable $W$ for a given fabrication floor.

\section{Results}
\label{sec:results}

This section substitutes the scaling laws from section~\ref{sec:methods} into the SNR expression (equation~\ref{eq:SNR}) and optimizes the result for different sample geometries. Throughout this section we assume pure additive noise; the effect of multiplicative baseline noise is incorporated in section~\ref{sec:phi}. The strategy in each case is to substitute the $W$-scaling of $\alpha_\mathrm{s}$ and $\alpha_\mathrm{c}$, factor out $W$-independent quantities, and reduce the optimization to a single dimensionless variable $u = \alpha_\mathrm{c} L$.

\begin{figure}[t]
\centering
\includegraphics[width=\linewidth]{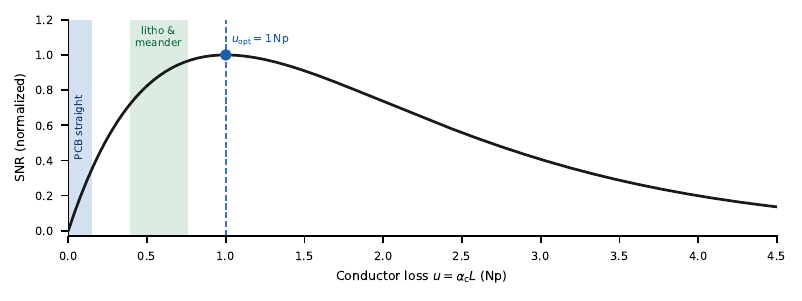}
\caption{Normalized SNR as a function of conductor loss $u = \alpha_\mathrm{c}L$. The $u\,e^{-u}$ curve (solid) is normalized to its maximum; the optimum at $1\,\mathrm{Np}$ (filled circle) is the Twyman--Lothian result under additive noise. Blue shading: region accessible to PCB straight-line instruments ($u \lesssim 0.38\,\mathrm{Np}$). Green shading: range achieved by published photolithographic designs ($u \approx 0.33$--$0.95\,\mathrm{Np}$); see figure~\ref{fig:fmr_scatter} for per-design values.}
\label{fig:fom}
\end{figure}

The simplest case is where the sample coats the entire CPW surface (a powder layer, spin-coated film, or nanoparticles from solution) and both $W$ and $L$ are free design parameters. From sections~\ref{sec:filling} and \ref{sec:loss}:
\begin{equation*}
  \alpha_\mathrm{s} \propto \frac{t}{W}, \qquad
  \alpha_\mathrm{c} \propto \frac{1}{W},
\end{equation*}
so their ratio $\alpha_\mathrm{s}/\alpha_\mathrm{c} \propto t$ is independent of $W$, as shown explicitly in Figure~\ref{fig:geometry}(c) across three decades of $W$ throughout the thin-sample regime. Narrowing the conductor concentrates the RF field (increasing $\eta$) and increases the current density (increasing ohmic loss) in exact proportion, so the two effects cancel. Writing $\alpha_\mathrm{s} = (\alpha_\mathrm{s}/\alpha_\mathrm{c})\,\alpha_\mathrm{c}$ and $u = \alpha_\mathrm{c} L$, the SNR becomes
\begin{equation*}
  \mathrm{SNR} \propto \frac{\alpha_\mathrm{s}}{\alpha_\mathrm{c}}
    \times u \times e^{-u}.
\end{equation*}
The first factor is $W$-independent. Maximizing over $L$ gives the Twyman--Lothian~\cite{TwymanLothian1933} result expressed as a CPW design rule under additive noise:
\begin{equation}
  \boxed{\alpha_\mathrm{c}\,L = 1\Np = 8.686\dB}
  \qquad\text{(full coverage, additive noise)}.
  \label{eq:optimal_regA}
\end{equation}
At this point the marginal gain in signal from a longer interaction path is exactly offset by the marginal penalty from exponential background attenuation. Figure~\ref{fig:fom} shows the $u\,e^{-u}$ optimization curve.

Because $W$ drops out of the optimization entirely, all values of $W$ yield the same SNR provided $L$ satisfies equation~\ref{eq:optimal_regA}. Since $\alpha_\mathrm{c} \propto 1/W$, the required length scales as $L_\mathrm{opt} \propto W$: narrow conductors require short CPWs; wide conductors require long ones. The design is therefore set by practical constraints---available sample area, minimum achievable $W$, or maximum practical $L$. When a specific sample area is fixed, the next question is how to route the CPW through it optimally.

\subsection{Fixed measurement area: meander geometry}
\label{sec:meander}

When the sample or apparatus fixes the available measurement area $w_\mathrm{s} \times l_\mathrm{s}$, routing multiple parallel passes through the sample maximises spatial coverage while preserving the $u\,e^{-u}$ SNR profile and its $1\,\mathrm{Np}$ optimum. The central practical question is whether the required conductor width is achievable with the available fabrication process. This analysis assumes the sample is smaller than a single meander run; for large-area extended films that span multiple passes, each run contributes additively to the total interaction length and the meander is equivalent to a straight CPW of the same total length $L$---a useful strategy when $L_\mathrm{opt}$ exceeds practical straight-line dimensions.

If the runs are spaced at pitch $P = \beta W$, where $\beta \geq 2$ is a geometry factor set by the minimum ground-plane clearance of the fabrication process ($\beta = 2$ is the dense-packing limit), the number of passes across the sample width is $M \approx w_\mathrm{s}/(\beta W)$, giving a total meander path length
\begin{equation}
  L = M \, l_\mathrm{s} \approx \frac{w_\mathrm{s}\, l_\mathrm{s}}{\beta W}
    \propto \frac{1}{W}.
  \label{eq:meander_L}
\end{equation}
This $L \propto 1/W$ constraint links $L$ and $W$ in a way distinct from the full-coverage case ($L$ free) and the single-pass case ($L$ fixed). In the thin-sample regime ($t/W \lesssim 0.1$), the per-unit-length coupling and conductor loss retain the same $W$-scaling as the full-coverage case, so the normalized SNR is $u\,e^{-u}$ with maximum at $\alpha_\mathrm{c} L = 1\,\mathrm{Np}$ (additive noise). The bulk-sample case ($t \gtrsim G$) shifts this optimum and is treated in section~\ref{sec:interpretation}. Unlike the full-coverage case, the meander constraint selects a \emph{unique} optimal conductor width. From equations~\ref{eq:Nsq} and \ref{eq:meander_L}, the aspect ratio of the meander is $L/W \approx w_\mathrm{s} l_\mathrm{s}/(\beta W^2)$. Setting $L/W$ equal to its optimum value (derived in section~\ref{sec:thinfilm}) gives
\begin{equation}
  W_\mathrm{opt} = \sqrt{\frac{A_\mathrm{c}\,\Rs\,w_\mathrm{s}\,l_\mathrm{s}}{\beta}},
  \qquad
  L_\mathrm{opt} = \frac{W_\mathrm{opt}}{A_\mathrm{c}\,\Rs}.
  \label{eq:meander_Wopt}
\end{equation}

For a 50\,$\Omega$ CPW the conformal-mapping condition links the minimum conductor width to the minimum slot: $W_\mathrm{min} \approx 12\,S_\mathrm{min}$ on RO4003C ($\varepsilon_r = 3.55$), $\approx 10.5\,S_\mathrm{min}$ on fused silica ($\varepsilon_r = 3.8$), and $\approx 2.4\,S_\mathrm{min}$ on sapphire ($\varepsilon_r = 9.4$). Table~\ref{tab:meander_examples} collects the implied meander thresholds and example outcomes for the three principal fabrication classes.

\begin{table}[t]
\caption{Meander design outcomes for three fabrication classes at $\beta = 3$ (PCB example uses $\beta = 2$). $S_\mathrm{min}$: minimum patternable slot; $W_\mathrm{min}$: implied minimum conductor width; $L_{s,\mathrm{min}}$: smallest square sample for which the meander optimum is accessible ($W_\mathrm{opt} \geq W_\mathrm{min}$). Example results use $W = W_\mathrm{opt}$.}
\label{tab:meander_examples}
\centering\small
\begin{tabular}{@{}lcccccc@{}}
\toprule
Fabrication & Metallization & $S_\mathrm{min}$ & $W_\mathrm{min}$ &
$L_{s,\mathrm{min}}$ & Example & Result \\
\midrule
PCB milling     & Cu / RO4003C      & 76\,$\mu$m & 890\,$\mu$m & 18\,mm
  & 25\,mm film, $\beta{=}2$
  & 10 passes, $u\approx1$\,Np \\[3pt]
Optical litho   & Au / fused silica &  5\,$\mu$m &  52\,$\mu$m & 490\,$\mu$m
  & 1\,mm sample
  & 3 passes, $u\approx1$\,Np \\[3pt]
E-beam          & Au / fused silica & 0.5\,$\mu$m & 5\,$\mu$m &  47\,$\mu$m
  & 100\,$\mu$m sample
  & 3 passes, $u\approx1$\,Np \\
\bottomrule
\end{tabular}
\end{table}

For optical and e-beam designs on fused silica the pass count is sample-size independent for square samples: $M \approx 1/\!\sqrt{\beta\,A_\mathrm{c}\Rs} \approx 3$ for $\beta = 3$ and 100\,nm Au. A 30\,$\mu$m\,$\times$\,30\,$\mu$m exfoliated crystal falls below the e-beam meander threshold ($W_\mathrm{opt} \approx 3.2\,\mu$m $< W_\mathrm{min}$) and illustrates the single-pass case that follows.

\subsection{Fabrication-limited designs: sample smaller than minimum conductor width}
\label{sec:fabrication_limited}

When the sample width is smaller than the minimum patternable conductor width ($w_\mathrm{s} \lesssim W_\mathrm{min}$), neither a meander nor a single-pass design can achieve the $1\,\mathrm{Np}$ optimum: the conductor always overfills the sample, reducing coupling without a corresponding reduction in conductor loss. The only path to the optimum is to switch to a fabrication process with a finer minimum slot, moving the design back into the meander regime of section~\ref{sec:meander}. The design choice reduces to a single comparison: compute $W_\mathrm{opt}^\mathrm{meander}$ from equation~\ref{eq:meander_Wopt} and compare it to $W_\mathrm{min}$. If $W_\mathrm{min} \leq W_\mathrm{opt}^\mathrm{meander}$ the meander is achievable and always preferred; otherwise a single pass at $W_\mathrm{target} = A_\mathrm{c}\Rs\,l_\mathrm{s}$ is the best available design. In all cases the target is $u_\mathrm{opt} = 1\,\mathrm{Np}$ under additive noise, with the fabrication floor determining how closely it can be approached.

\section{Discussion}
\label{sec:discussion}

The results of section~\ref{sec:results} assume pure additive noise. In every regime, the shared $W$-dependence of $\alpha_\mathrm{s}$ and $\alpha_\mathrm{c}$ allows the SNR to be factored into a $W$-independent prefactor and a function of $u = \alpha_\mathrm{c}L$ alone, giving the central design equation
\begin{equation}
  \boxed{u_\mathrm{opt} = \frac{1}{1-\varphi}},
  \label{eq:u_opt_combined}
\end{equation}
where $\varphi \in [0,1]$ interpolates between pure additive and pure multiplicative noise. The parameter is discussed in section~\ref{sec:phi}; the geometric analysis of section~\ref{sec:results} is independent of $\varphi$ and applies at any noise regime.

\subsection{Noise regime and the $\varphi$ parameterization}
\label{sec:phi}

The results of section~\ref{sec:results} assume that receiver noise dominates. In practice, the dominant noise source in VNA-based spectroscopy is often not the receiver floor but \emph{multiplicative baseline structure}: standing waves from connector reflections, impedance mismatches, and residual calibration errors produce a fractional uncertainty $\varepsilon$ in $\SII$ that scales with the transmitted signal. In this limit, $\mathrm{SNR} \propto \alpha_\mathrm{s}L/\varepsilon$, which grows monotonically with path length and has no finite optimum. The additive-noise $1\Np$ result is therefore a lower bound on useful conductor loss, not a universal design target.

For field-swept EPR and FMR spectroscopy, however, the field-subtraction procedure described in the introduction substantially suppresses these multiplicative contributions, since field-independent standing waves and connector reflections cancel in $\Delta S_{21} = S_{21}(B_0) - S_{21}(B_\mathrm{ref})$. The effective $\varphi$ is therefore close to zero for this class of measurement, and the $1\Np$ target applies directly rather than serving only as a lower bound.

Real measurements mix both noise sources. Writing the total noise variance as
\begin{equation*}
  \sigma^2 = (1 - \varphi)\,\sigma_\mathrm{a}^2
           + \varphi\,(\varepsilon\,|\SII|)^2,
\end{equation*}
where $\varphi \in [0,1]$ is a continuous mixing fraction parameterizing the relative weight of multiplicative noise. Maximizing the resulting SNR expression yields equation~\ref{eq:u_opt_combined}, which connects the geometric analysis of section~\ref{sec:results} to the noise regime of the measurement apparatus. Rothman, Crouch, and Ingle~\cite{Rothman1975} classified the dominant noise sources in optical spectrophotometry by how the standard deviation $s_T$ of the transmittance $T$ scales with $T$: constant $s_T$ (Category~1, detector/readout noise), $s_T \propto \sqrt{T}$ (Category~2, photon shot noise), and $s_T \propto T$ (Category~3, source flicker). In microwave transmission via VNA, the categories with natural analogs are Category~1, which maps to the VNA receiver Johnson floor, and Category~3, which maps to baseline ripple from standing waves, impedance mismatches, and residual calibration errors. Category~2 has no natural microwave analog: at room temperature, $k_\mathrm{B}T \gg h\nu$ throughout the GHz band, so microwave detection is not photon-shot-noise limited and the receiver thermal noise instead behaves as Category~1. The correspondence between the realized optical categories and their CPW analogs, including the conversion $1\;\mathrm{AU} = 2.303\;\mathrm{Np}$, is summarized in table~\ref{tab:noise_categories}. Equation~\ref{eq:u_opt_combined} interpolates between these two limits as $\varphi$ runs from $0$ (Cat~1) to $1$ (Cat~3).

\begin{table}[b]
\caption{Rothman--Crouch--Ingle noise categories~\cite{Rothman1975} with natural microwave analogs and the corresponding values of the mixing parameter $\varphi$. Category~2 (photon shot noise, $s_T \propto \sqrt{T}$, optimum at $\alpha L = 2\;\mathrm{Np}$) is omitted because microwave detection is not photon-shot-noise limited at room temperature. Optimal values are related by $A \times \ln 10 = \alpha L$, with $1\;\mathrm{AU} = 2.303\;\mathrm{Np}$.}
\label{tab:noise_categories}
\centering\small
\begin{tabular}{@{}clllll cc@{}}
\toprule
& & & & &
  & \multicolumn{2}{c}{Optimal value} \\
\cmidrule(l){7-8}
Cat. & $\varphi$ & $s_T$ scaling & Optical source & CPW analog
  & & $A$ (AU) & $\alpha L$ (Np) \\
\midrule
1 & $0$     & Constant    & Detector/readout & VNA receiver floor & & 0.43        & 1           \\
3 & $\to 1$ & $\propto T$ & Source flicker   & Baseline ripple    & & $\to\infty$ & $\to\infty$ \\
\bottomrule
\end{tabular}
\end{table}

This parameterization accounts for two observations. First, practitioners use longer transmission lines than the $1\,\mathrm{Np}$ rule predicts \cite{Clauss2013,Boudalis2021}: baseline ripple shifts the effective optimum upward, and the SNR is relatively insensitive to operation above $u = 1\Np$ when $\varphi > 0$. Second, PCB-fabricated FMR instruments (figure~\ref{fig:fmr_scatter}) operate well below $1\,\mathrm{Np}$ yet still yield useful results: in the multiplicative limit ($\varphi \to 1$) the SNR grows monotonically with path length and there is no penalty for low $u$. Movement toward $u_\mathrm{opt}$ improves sensitivity multiplicatively in all cases.

\subsection{Thin-film metallization and the optimal aspect ratio}
\label{sec:thinfilm}

Combining equations~\ref{eq:Nsq} and \ref{eq:u_opt_combined}, the optimization target $\alpha_\mathrm{c} L = u_\mathrm{opt}$ requires a conductor with aspect ratio
\begin{equation*}
  \frac{L}{W}\bigg|_\mathrm{opt} = \frac{u_\mathrm{opt}}{A_\mathrm{c}\,\Rs}
    = \frac{1}{(1-\varphi)\,A_\mathrm{c}\,\Rs}.
\end{equation*}
The product $\Rs \cdot (L/W)$ is precisely the end-to-end sheet resistance of the conductor strip --- the same quantity measured in a van der Pauw or Hall bar resistivity structure. The CPW optimization therefore reduces to a familiar thin-film target: achieve a specific sheet resistance $\Rs(L/W) = u_\mathrm{opt}/A_\mathrm{c}$ in the center conductor. A higher $\Rs$ means a shorter, wider conductor suffices.

The surface resistance depends on the metal thickness $t_\mathrm{m}$ relative to the skin depth $\delta_\mathrm{s} = \sqrt{1/(\pi f \mu_0 \sigma)}$. In the bulk regime ($t_\mathrm{m} \gg \delta_\mathrm{s}$), $\Rs = \sqrt{\pi f \mu_0/\sigma}$ grows as $\sqrt{f}$, so the optimal aspect ratio is frequency-dependent and a broadband CPW cannot be simultaneously optimal at all frequencies. In the thin-film regime ($t_\mathrm{m} \ll \delta_\mathrm{s}$), $\Rs = 1/(\sigma t_\mathrm{m})$ is frequency-independent, so the optimal design point is constant across the full measurement bandwidth. For gold at 10~\GHz, $\delta_\mathrm{s} \approx 790$~nm, so lithographic CPWs with 50--300~nm metallization operate in the thin-film regime.

The thin-film $\Rs$ is also substantially larger than its bulk counterpart. For 100~nm Au at 10~\GHz: $\Rs^\mathrm{thin} \approx 0.24\;\Omega/\square$ compared with $\Rs^\mathrm{bulk} \approx 0.037\;\Omega/\square$ for PCB-grade copper (conductivity $\approx 50\%$ of ideal, typical for electrodeposited Cu)---a factor of six to seven. The target aspect ratio is correspondingly reduced: for 100~nm Au on fused silica at $\Zo = 50\;\Omega$ and pure additive noise, $(L/W)_\mathrm{opt} \approx 30$, achievable in a single straight run. With bulk copper, the same target requires $L/W \approx 200$, necessitating a meander geometry that introduces fabrication complexity. Using the bulk formula for a thin-film CPW would overestimate the target aspect ratio by the same factor, producing a design that is far too long and far too lossy. The crossover between the bulk and thin-film regimes is smooth and is described by the interpolation \cite{Pozar2012,Heinrich1993}
\begin{equation}
  \Rs = \frac{1}{\sigma\, \delta_\mathrm{s}}
        \coth\!\left(\frac{t_\mathrm{m}}{\delta_\mathrm{s}}\right),
  \label{eq:Rs_interp}
\end{equation}
which recovers the thin-film and bulk limits as $t_\mathrm{m}/\delta_\mathrm{s} \to 0$ and $\to \infty$ respectively. This interpolation is used in the design tool described in section~\ref{sec:tool}.

Increasing $\Rs$ also reduces $W_\mathrm{opt}^\mathrm{meander} \propto \sqrt{\Rs}$, demanding finer lithography to access the meander regime. For a 30~$\mu$m $\times$ 30~$\mu$m sample, switching from bulk copper (target aspect ratio $\approx 200$) to 100~nm Au (target aspect ratio $\approx 30$) moves $W_\mathrm{opt}^\mathrm{meander}$ from approximately 1.2~$\mu$m (below the e-beam fabrication floor) to 3~$\mu$m (just accessible by e-beam). Thin-film metallization therefore has the greatest impact when combined with high-resolution patterning; reducing $t_\mathrm{m}$ without improving $W_\mathrm{min}$ only trades one fabrication limit for another.

\subsection{Benchmarking: transmission-line FMR designs}
\label{sec:fmr_benchmarks}
\begin{figure}[t]
\centering
\includegraphics[width=0.95\columnwidth]{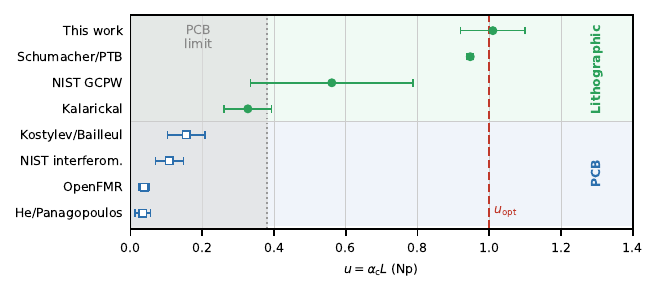}
\vspace{-5mm}
\caption{Conductor loss $u \pm 1\sigma$ for seven published broadband FMR transmission-line designs~\cite{Edwards2017,Glowinski2014,HePanagopoulos2016,OpenFMR2025,Kalarickal2006,Schoen2015,Liebing2011}, plus this work, ordered by increasing $u$. Open squares: PCB-fabricated designs (PCB-grade bulk Cu). Filled circles: photolithographically patterned designs (thin-film Au). Dashed line: $u_\mathrm{opt} = 1\Np$ under additive noise. Grey shading: region accessible to PCB fabrication ($W \geq 200\,\mu$m, $L \leq 15$~mm, PCB-grade bulk Cu on RO4003C at 10~GHz); the dotted boundary at $u \approx 0.38\Np$ is a ceiling imposed by the minimum patternable conductor width. All four PCB designs lie below this ceiling; the three previous lithographic designs span the gap to the optimum, with Schumacher/PTB essentially reaching it. Full per-design calculation details are given in the Supplementary Material.}
\label{fig:fmr_scatter}
\end{figure}

Figure~\ref{fig:fmr_scatter} benchmarks seven published broadband FMR transmission-line designs against the $u = 1\Np$ optimum, spanning conductor widths from 64~$\mu$m to 450~$\mu$m. The conductor loss $u$ for each was estimated using the model of section~\ref{sec:loss} with Monte Carlo uncertainty propagation over all stated and assumed geometric parameters; full per-design detail is provided in the Supplementary Material.  Across the seven designs, $u$ ranges from $0.03$ to $0.95\,\mathrm{Np}$, corresponding to 9\%--100\% of the optimal SNR.

The results reveal a sharp division between the two fabrication classes. All four PCB designs operate at $u \lesssim 0.16\Np$---between 9\% and 36\% of the optimal SNR---despite spanning a fourfold range of conductor widths (117--450~$\mu$m) and two CPW topologies. The two grounded-CPW designs (OpenFMR, He/Panagopoulos) sit at $u \approx 0.04\,\text{Np}$ despite their narrower $W$ (117 and 180\,$\mu$m): their wide slots ($S/W \gtrsim 0.6$) reduce the slot-edge contribution to $\alpha_c$, the dominant conductor loss term in narrow-gap CPWs. For PCB-grade bulk Cu at any of these widths, the required $L_\mathrm{opt} = W/(A_\mathrm{c}\Rs)$ ranges from 23 to 88~mm---exceeding practical straight-line PCB dimensions.

The three photolithographically patterned designs reach $u = 0.33$--$0.95\Np$ (64\%--100\% of the optimal SNR) despite operating on smaller samples. The smaller $W$ (64--100~$\mu$m) increases $\Nsq$ and the stronger field confinement increases $A_\mathrm{c}$; the elevated thin-film $\Rs$ of pure-Au designs (3--7$\times$ bulk Cu at 10~GHz) reduces the required $L_\mathrm{opt}$ to a few millimeters, making the optimum accessible in a single straight pass. Schumacher/PTB at $u = 0.947\Np$ reaches the additive-noise optimum to within 1\% as an unintentional consequence of choosing thin Au for magnetic adhesion reasons. The present design at $W = 100\,\mu$m, $L = 2.96$~mm on a low-dielectric-loss fused-silica substrate also reaches the optimum, deliberately.

The sensitivity workarounds adopted by the PCB-fabricated designs are consistent with this picture.  The NIST interferometer \cite{Edwards2017} reports $|\Delta S_{21}| \approx -50\,\mathrm{dB}$ for a typical Py film and employs a Michelson microwave interferometer providing $+35\,\mathrm{dB}$ of background suppression to recover a usable SNR. OpenFMR~\cite{OpenFMR2025} uses lock-in field modulation to detect 1~nm CoFeB films.  Both are adaptations to a fabrication ceiling that makes the signal budget insufficient for direct VNA detection: the $u\,e^{-u}$ framework quantifies the SNR shortfall that these workarounds offset.

\subsection{Physical interpretation and model consistency}
\label{sec:interpretation}

The benchmarking results in section~\ref{sec:fmr_benchmarks} rank designs by conductor loss, but the connection to measured signal levels is not self-evident.  The link is established by the radiative damping formalism of Schoen \emph{et al.}~\cite{Schoen2015}, who showed that the radiative contribution to the FMR linewidth scales as $\alpha_\mathrm{rad} \propto M_\mathrm{s}^2\,\delta\,l / (f^2\,W)$. Because $|\Delta S_{21}|$ at resonance is proportional to $\alpha_\mathrm{rad}$ via Faraday's law~\cite{Edwards2017}, it carries the same $l/W$ dependence captured by $u = A_\mathrm{c}\Rs(L/W)$ in the $u\,e^{-u}$ framework. Consequently, the ratio of SNRs between any two designs equals the ratio of their absolute $|\Delta S_{21}|$ values, and the $\sim$12~dB amplitude improvement predicted between the NIST interferometer ($u = 0.108\,\mathrm{Np}$) and the present design ($u = 1.01\,\mathrm{Np}$) accounts for roughly one-third of the $+35\,\mathrm{dB}$ of interferometric suppression the NIST instrument employs to recover a usable SNR.

The Beer--Lambert structure of the optimization persists across the full range of sample thicknesses, though the optimal conductor loss depends on the geometric constraint on $L$ once $\eta$ saturates. In the thin-film limit ($t \ll W$) the shared $1/W$ scaling of $\alpha_\mathrm{s}$ and $\alpha_\mathrm{c}$ gives $u_\mathrm{opt} = 1\,\mathrm{Np}$ regardless of how $L$ is constrained. In the bulk limit ($t \gtrsim G$), $\eta$ saturates at a constant $\eta_\mathrm{bulk}$ and $\alpha_\mathrm{s}$ becomes $W$-independent: a free-$L$ design retains $u_\mathrm{opt} = 1\,\mathrm{Np}$ but with the $W$-degeneracy broken (peak SNR scales as $\eta_\mathrm{bulk}\times W$, so $W$ is chosen as wide as practical constraints allow), while a meander over fixed area $w_s\times l_s$ has $L\propto 1/W$ which combined with $W$-independent $\alpha_\mathrm{s}$ gives $\alpha_\mathrm{s}L\propto\sqrt{u}$ and shifts the optimum to $u_\mathrm{opt} = 1/2\,\mathrm{Np}$ with $W_\mathrm{opt}$ a factor of $\sqrt{2}$ wider than the thin-film prescription. None of the benchmarked instruments in figure~\ref{fig:fmr_scatter} operate in the bulk regime, so the $1/2\,\mathrm{Np}$ result stands as a forward-looking design rule rather than a fit to existing data.

\subsection{Design tool}
\label{sec:tool}

An interactive web-based tool, \textsc{NpTuner}, implementing the full analytical model of sections~\ref{sec:methods} and \ref{sec:results} is freely available; source code is hosted at \url{\nptunerurl} with a link to the running web application. The user specifies sample geometry (thickness, lateral dimensions, coverage type), substrate material, metallization, and an estimate of $\varphi$ for the measurement apparatus. The tool evaluates the filling factor, conductor and dielectric loss, and SNR across the full range of conductor widths, identifies the optimal design point using equation~\ref{eq:u_opt_combined}, and plots the $u\,e^{-u}$ curve with the $\varphi$-shifted operating point marked. For cases where $t/W > 0.1$ and the thin-sample approximation breaks down, a full numerical conformal-mapping integral is available. The conformal-mapping integral (equation~\ref{eq:eta_def}) agrees with finite-element solutions to within 15\% for $t/W < 0.1$ \cite{Simons2001}, covering thin-film and nanoparticle applications.

\section{Conclusions}
\label{sec:conclusions}

The shared $1/W$ scaling of sample coupling and conductor loss in CPW transmission spectroscopy reduces the design optimisation for fixed-volume samples to a universal one-parameter problem.  The central result is $u_\mathrm{opt} = 1/(1-\varphi)$, where $u = \alpha_\mathrm{c}L$ is the conductor loss in nepers and $\varphi$ encodes the noise regime.  The $1\,\mathrm{Np}$ Twyman--Lothian result~\cite{TwymanLothian1933} is the $\varphi = 0$ special case; since field-subtraction in EPR and FMR measurements suppresses multiplicative baseline noise, this is the directly actionable design target for that class of measurement. The optimal conductor loss is independent of sample properties, operating frequency, and system losses---it depends only on what noise dominates.  For a fixed sample area, the meander geometry is preferred whenever the fabrication floor satisfies $W_\mathrm{min} \leq W_\mathrm{opt}^\mathrm{meander} = \sqrt{A_\mathrm{c}\Rs w_\mathrm{s} l_\mathrm{s}/\beta}$; a single-pass design at $u_\mathrm{target} = A_\mathrm{c}\Rs\,l_\mathrm{s}/W_\mathrm{target}$ is the best achievable otherwise, with each improvement in lithographic resolution moving the design closer to the $1\,\mathrm{Np}$ optimum.

Benchmarking against seven published broadband FMR instruments confirms a sharp division between fabrication classes.  PCB-milled instruments are bounded by a conductor-loss ceiling well below the optimum: PCB-grade bulk-Cu $\Rs$ requires $L_\mathrm{opt} \approx 174$~mm at the minimum slot-constrained width of $W \approx 890\,\mu$m on RO4003C, far exceeding practical straight-line dimensions. A meander over a 25~mm sample accommodates 10 passes at $\beta = 2$ and reaches the additive-noise optimum $u \approx 1\,\mathrm{Np}$. This is substantially better than the 9--36\% of optimal SNR achieved by current straight-line PCB designs, though the benefit diminishes rapidly for smaller samples below the meander threshold $L_{s,\mathrm{min}} \approx 18$~mm. Photolithographic designs reach $u = 0.33$--$1.0\,\mathrm{Np}$ through narrower $W$ and the elevated $\Rs$ of thin-film metallization, which also renders the optimal aspect ratio frequency-independent across the full broadband measurement range. The sensitivity-enhancement strategies in current use---Michelson interferometry~\cite{Edwards2017}, lock-in field modulation~\cite{OpenFMR2025}, and extended signal averaging---are rational responses to the PCB fabrication ceiling, and their convergent adoption is consistent with the signal-level predictions of the $u\,e^{-u}$ framework.

The $1/W$ scaling that underlies these results holds throughout the thin-sample regime ($t \ll W$) and wherever the logarithmic correction in the Pucel conductor-loss formula is slowly varying.  When sample thickness becomes comparable to $W$, when impedance varies significantly across the design space, or when the coverage fraction changes in a non-trivial way with $W$, the analytical optima serve as starting points for a numerical optimization using the full expressions in the \textsc{NpTuner} design tool available at \url{\nptunerurl}.

\ack{The authors thank David Chuss (Villanova University) and Lloyd Engel (National High Magnetic Field Laboratory) for valuable discussions.}

\section*{Conflict of interest}
The authors declare that they have no known competing financial interests or personal relationships that could have appeared to influence the work reported in this paper.

\funding{This work was supported by the National Science Foundation under CAREER Award No.\ DMR-1943389.}

\roles{SD: Conceptualization, Methodology, Writing -- original draft, Software. AS: Investigation, Validation.}

\data{The data that support the findings of this study are openly available. The interactive design tool \textsc{NpTuner} is available at \url{\nptunerurl}.}


\end{document}